\documentclass[twocolumn,apl, superscriptaddress,longbibliography]{revtex4-2}

\usepackage{graphicx}
\usepackage{color}
\usepackage{amsfonts}
\usepackage[version=4]{mhchem}
\usepackage{isomath}
\usepackage{amsmath}
\usepackage{amsthm}
\usepackage{amsfonts}
\usepackage{amssymb}
\usepackage{fdsymbol}
\usepackage{braket}
\usepackage{gensymb}

\let\oldAA\AA
\renewcommand{\AA}{\text{\normalfont\oldAA}}

\begin{document}

\title{Band offsets and stability of WSe$_2$/RuCl$_3$ van der Waals charge-transfer contacts}
\author{Thomas S.  Nielsen}
\author{Edvard Solbrekken}
\author{Christian Overby}
\author{Christian V-B. Fokdal}
\author{Alfred J. H. Jones}
\author{Zhihao Jiang}
\author{Chakradhar Sahoo}
\affiliation{Department of Physics and Astronomy, Aarhus University, 8000 Aarhus C, Denmark}
\author{Kenji Watanabe}
\affiliation{Research Center for Electronic and Optical Materials, National Institute for Materials Science, 1-1 Namiki, Tsukuba 305-0044, Japan}
\author{Takashi Taniguchi}
\affiliation{Research Center for Materials Nanoarchitectonics, National Institute for Materials Science, 1-1 Namiki, Tsukuba 305-0044, Japan}
\author{Søren Ulstrup}
\email{ulstrup@phys.au.dk}
\affiliation{Department of Physics and Astronomy, Aarhus University, 8000 Aarhus C, Denmark}

\date{\today}

\begin{abstract}

The layered Mott insulator $\alpha$-RuCl$_3$ induces degenerate hole-doping in two-dimensional semiconductors due to its large electron affinity,  making it a promising charge-transfer material for establishing ohmic contacts in electronic devices.  In order to assess the applicability and guide the design of devices incorporating RuCl$_3$ it is critical to determine the electronic structure and robustness of the band offsets that underpin the transport properties of semiconductors in contact with RuCl$_3$. Here, we apply micro-focused angle-resolved photoemission spectroscopy to determine the electronic structure of single-layer WSe$_2$ contacted to RuCl$_3$ on hexagonal boron nitride substrates. We find that formation of a functioning WSe$_2$/RuCl$_3$ contact leads to a valence band shift of $(0.68 \pm 0.05)$ eV towards the Fermi energy in WSe$_2$. The charge transfer effect is challenging to observe as it depends sensitively on fabrication conditions such as solvent exposure, quality of interface encapsulation and heating of RuCl$_3$, imposing strict requirements on device design to attain high-quality contacts.
\end{abstract}

\maketitle

Two-dimensional (2D) transition metal dichalcogenides (TMDs) with composition MX$_2$, where M$=\{$Mo,W$\}$ and X$=\{$S,Se$\}$, are attractive materials for ultra-scaled transistors and optoelectronic devices as they are direct gap semiconductors with large on/off switching ratios and low power dissipation  \cite{Radisavljevic:2011,Ovchinnikov:2014,Saptarshi:2021, Jayachandran:2024,wuSynthesisModulationApplication2024}. However, the salient transport properties of TMDs directly contacted to bulk metallic electrodes are impeded by contact resistance,  which emerges from a difference between electrode work function and semiconductor electron affinity,  leading to a Schottky barrier \cite{Allain:2015,Schulman:2018,wangMakingCleanElectrical2022,liuFermiLevelPinning2022}.  Numerous strategies have been envisioned to solve this problem,  such as using graphene edge contacts \cite{Marcos:2016,Schneider:2024},  lateral 1T-1H TMD structure engineering \cite{Kappera:2014},  molecular doping \cite{Lingming:2014} as well as hybridization of TMD electronic bands with those of metals and semimetals \cite{Dendzik:2017,Shen:2021,Weisheng:2023}.  An alternative strategy applies van der Waals materials with a large electron affinity as charge-transfer contacts in heterostructures with TMD semiconductors, leading to a strong hole doping of the lower electron affinity semiconductor \cite{Chanana:2016,Likuan:2024}. This is an attractive approach because clean van der Waals contacts are less prone to induce in-gap states and Fermi level pinning and thus uncontrollable Schottky barriers \cite{Ulstrupim:2019,Likuan:2024}.  

\begin{figure}[t!]
\begin{center}
\includegraphics[scale=1]{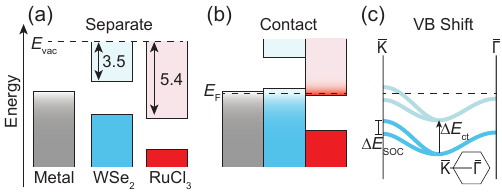}
\caption{Band alignments in a metal/WSe$_2$/RuCl$_3$ junction.  (a) Diagram of valence and conduction band alignments in the separate materials referenced to a common vacuum level ($E_{\mathrm{vac}}$).  Electron affinities are illustrated by double-headed arrows and given in units of eV \cite{Guo:2016,Klaproth:2022}.  (b) Band alignments after contact formation and equilibration of Fermi level ($E_{\mathrm{F}}$).  Grey,  blue and red color gradients illustrate carrier filling in the two scenarios in (a)-(b). 
(c) Schematic of SL WSe$_2$ valence band before (blue) and after (light-blue) contact formation.  Charge transfer leads to an energy shift ($\mathrm{\Delta}E_{\mathrm{ct}}$) placing the spin-orbit coupling split ($\mathrm{\Delta}E_{\mathrm{SOC}}$) valence band states around $E_{\mathrm{\mathrm{F}}}$.  The bands are illustrated along the $\bar{\mathrm{\Gamma}}$-$\bar{\mathrm{K}}$ direction indicated by a grey bar in the BZ inset.}
\label{Fig:1}
\end{center}
\end{figure}

The Mott insulating and quantum spin liquid candidate material $\alpha$-RuCl$_3$ (referred to as RuCl$_3$ from here) is a particularly promising charge-transfer contact material due to its large electron affinity of 5.4 eV \cite{Pollini:1996,polliniPhotoemissionStudyElectronic1994,zhouEvidenceChargeTransfer2019a,mashhadiSpinSplitBandHybridization2019a,Klaproth:2022,rossiDirectVisualizationCharge2023}.  Contacting RuCl$_3$ with single-layer (SL) WSe$_2$ leads to an especially interesting situation as the electron affinity of WSe$_2$ is merely 3.5 eV \cite{Guo:2016},  which results in a type-III band alignment as illustrated in Fig.  \ref{Fig:1}(a).  Upon forming a metal/WSe$_2$/RuCl$_3$ junction,  electrons transfer to RuCl$_3$ from WSe$_2$,  placing the Fermi level, defined by the metal contact, below the valence band maximum (VBM) of  WSe$_2$,  as shown in  Fig.  \ref{Fig:1}(b).  This junction geometry has led to the observation of a hole doping density reaching 3$\times$10$^{13}$cm$^{-2}$ in SL WSe$_2$ and thereby formation of an ohmic contact \cite{xieLowResistanceContact2024}.  From a band structure perspective,  the spin-orbit coupling (SOC) split and light hole-type bands around the $\bar{\mathrm{K}}$-valleys in SL WSe$_2$ are shifted into the regime of low-energy charge carrier transport as illustrated in Fig.  \ref{Fig:1}(c)   \cite{choModulationDopingSingleLayer2022,aroraEngineeringInterfacialCharge2025,wangModulationDopingTwoDimensional2020}.  This has resulted in hole mobilities of 80000 cm$^2$V$^{-1}$s$^{-1}$ and tunable hole densities that enable control of correlated phases in quantum devices based on single- and twisted bi-layer WSe$_2$ \cite{packChargetransferContactsMeasurement2024,guoSuperconductivity50degTwisted2025,guoAngleEvolutionSuperconducting2026}.

Spectroscopic measurements of energy- and momentum-resolved band structures of SL WSe$_2$/RuCl$_3$ contacts are lacking,  leaving open questions about the magnitude of charge-transfer induced energy shifts,  $\mathrm{\Delta}E_{\mathrm{ct}}$, achievable in the WSe$_2$ layer, as well as potential hybridization of SL WSe$_2$ valence bands with the valence and conduction band states in the type-III aligned RuCl$_3$. Furthermore, the homogeneity and stability of RuCl$_3$ in such a contact has not been explored,  despite RuCl$_3$ being sensitive to heating cycles and organic solvents present in typical van der Waals heterostructure fabrication approaches \cite{breitnerThermalDecompositionKitaev2023}.  A clean and high-quality interface is crucial for efficient charge transfer \cite{sternbachQuenchedExcitonsWSe22023,zhengUltrafastHighlyMobile2025}. In this Letter,  we address these open questions by first benchmarking the electronic structure of a bulk RuCl$_3$ crystal and then measuring the band dispersion of a SL WSe$_2$/RuCl$_3$ heterostructure using angle-resolved photoemission spectroscopy with micrometer spatial resolution (microARPES).

The experiments were performed at the microARPES end-station on the AU-SGM4 beamline at the ASTRID2 synchrotron light source at Aarhus University, Denmark  \cite{jonesSpatialAngleresolvedPhotoemission2025}.  Bulk RuCl$_3$ crystals (HQ Graphene) were cleaved at room temperature at a pressure below 10$^{-8}$ mbar, before being transferred into the microARPES ultra-high vacuum (UHV) chamber.  WSe$_2$/RuCl$_3$ heterostructures supported on hexagonal boron nitride (hBN) on gold-coated SiO$_2$/Si wafers were fabricated using mechanical exfoliation and the dry transfer method \cite{caldwell2010technique,Sahoo:2025}.  Exfoliated flakes of SL WSe$_2$, RuCl$_3$ and hBN with thicknesses of tens of nanometers were successively picked up using polycarbonate (PC) film on a polydimethylsiloxane (PDMS) stamp.  The WSe$_2$/RuCl$_3$/hBN stacks were deposited on the gold-coated substrate by melting the PC film at 180$^{\circ}$C for a few minutes. Polymers were then removed using chloroform, acetone, and isopropanol to obtain a clean, exposed surface for microARPES measurements. All fabrication steps involving RuCl$_3$ were performed in a protected N$_2$ atmosphere with O$_2$ and H$_2$O contents less than 1 ppm.  The sample was then transferred via a vacuum suitcase into the microARPES chamber,  without further heating steps.  The microARPES measurements on bulk RuCl$_3$ and heterostructures were performed at room temperature and at a pressure below $ 2 \times 10^{-10}$ mbar.  The energy-,  momentum- and spatial-resolution were better than 35 meV,  0.01 \AA$^{-1}$ and 4 µm,  respectively.

\begin{figure*}[t!]
\includegraphics[scale=1]{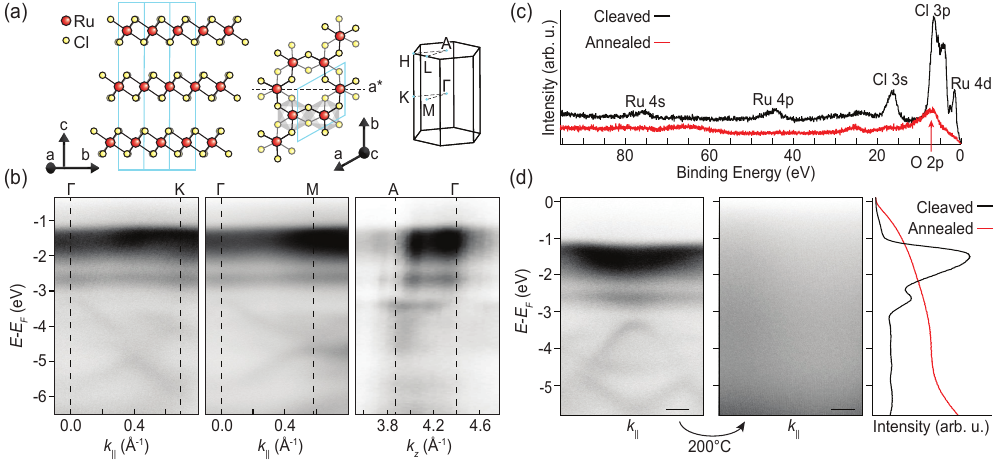}
\caption{Electronic structure of bulk RuCl$_3$ and impact of heating. (a) Crystal structure of RuCl$_3$ in the $P3_112$ space group with black arrows showing the crystallographic axes. Left is a side view along the a* direction of the 3 layers in the unit cell. Middle is a top view with the hexagonal unit cell indicated in blue and the RuCl$_6$ octahedra in gray. Right is the hexagonal bulk BZ. (b) ARPES spectra along $\mathrm{\Gamma}$-$\mathrm{K}$ and $\mathrm{\Gamma}$-$\mathrm{M}$ obtained at $h\nu=63$ eV, corresponding to the $\mathrm{\Gamma}$-K-M plane, and along $\mathrm{A}$-$\mathrm{\Gamma}$ by varying $h\nu$ between $35$-$80$ eV. An inner potential of 18 eV is used for the conversion to $k_z$ \cite{wangEvidenceWeylFermions2021}. (c) Core level spectra of bulk RuCl$_3$ before and after heating to $200^{\circ}$C measured at $h\nu=145$ eV.  (d) Valence band spectra from the same spot on the sample before and after heating to $200^{\circ}$C.  Momentum-integrated EDCs for each spectrum are presented to the right.  The spectra were obtained at $h\nu=55$ eV.  The scale bar is $0.2$ $\AA^{-1}$.}
\label{Fig:2}
\end{figure*}

We start with a characterization of the bulk electronic structure of an \textit{in situ} cleaved RuCl$_3$ flake from the same crystal we used to exfoliate flakes for the heterostructures,  providing a point of comparison for our heterostructure ARPES data presented later.  The crystal structure of RuCl$_3$ consists of layers of a hexagonal lattice of Ru atoms in edge sharing RuCl$_6$ octahedra,  which is visualized in Fig. \ref{Fig:2}(a) along with its bulk Brillouin zone (BZ) \cite{daiCrystalStructureReconstruction2020}.  ARPES spectra along $\mathrm{\Gamma}-\mathrm{K}$, $\mathrm{\Gamma}-\mathrm{M}$ and $\mathrm{A}-\mathrm{\Gamma}$ high symmetry directions of RuCl$_3$ are shown in Fig. \ref{Fig:2}(b).  Two flat bands are observed to be centered at -1.5 eV and -2.7 eV, which we identify as $t_{\mathrm{2g}}$ states deriving from the Ru 4d orbitals. At energies below -3 eV, we observe faint bands that are highly dispersive with $k_{\parallel}$ but relatively flat with $k_z$, which are characteristic of the Cl 3p states in the material \cite{sinnElectronicStructureKitaev2016,samantaElectronicStructuresKitaev2023a}. 

The ARPES results correspond well to previous studies and calculations of RuCl$_3$ in the $P3_112$ space group \cite{zhouAngleresolvedPhotoemissionStudy2016,sinnElectronicStructureKitaev2016,koitzschEffDescriptionHoneycomb2016,wangEvidenceWeylFermions2021}, ruling out any potential mixing of monoclinic $C2/m$ phases that can emerge due to stacking faults \cite{kimStructuralTransitionMagnetic2024,wangEvidenceWeylFermions2021,caoLowtemperatureCrystalMagnetic2016}.  The observed ARPES dispersion is consistent with a Mott insulating state, as the topmost valence band is a lower Hubbard band of the $t_{\mathrm{2g}}$ states with an effective angular momentum $J_{\textnormal{eff}}=1/2$ due to SOC \cite{plumbRuCl3Spinorbit2014,kimQuasimolecularBandInsulator2016,samantaElectronicStructuresKitaev2023a}.

Understanding the impact of heating RuCl$_3$ up to 200$^{\circ}$C is relevant both for the melting of polymers during heterostructure fabrication and for the final cleaning of heterostructures via vacuum annealing prior to ARPES or transport experiments.  We therefore exposed the RuCl$_3$ flake to a temperature of 200$^{\circ}$C in a pressure better than 10$^{-8}$ mbar for 1.5 hours.  In Fig.  \ref{Fig:2}(c),  we present core level spectra of as-cleaved and heated RuCl$_3$.  In the as-cleaved sample we observe several peaks between 0-20 eV, which correspond to Ru 4d, Cl 3p and Cl 3s valence states \cite{Pollini:1996}.  Ru 4s and Ru 4p core levels are identified at binding energies of 76.0 and 44.6 eV \cite{polliniPhotoemissionStudyElectronic1994}.  After heating,  the core level spectrum from the same spot on the same flake has dramatically changed with no signs of the Ru or Cl peaks observed in the as-cleaved case.  A broad feature is instead observed to extend from 5-10 eV, which we interpret as an oxygen 2p signal \cite{COX1986360}.  

To further elucidate the situation,  we compare ARPES spectra and momentum-integrated energy distribution curves (EDCs) of the as-cleaved and heated sample in Fig. \ref{Fig:2}(d).  After heating, the ARPES spectrum merely represents a uniform background with intensity extending to the Fermi level.  Ru and Cl derived bands are completely absent, which points towards degradation of RuCl$_3$.  This could possibly transpire through dechlorination and partial oxidation towards the metallic RuO$_2$ \cite{breitnerThermalDecompositionKitaev2023,newkirkThermalDecompositionRhodium1968,Pollini:1996},  which would explain the oxygen core level and the intensity at the Fermi level.  Thus, heating of the heterostructure should be kept to a minimum and ideally avoided to maintain RuCl$_3$  in a pristine state.

\begin{figure*}[t!]
\includegraphics[scale=1]{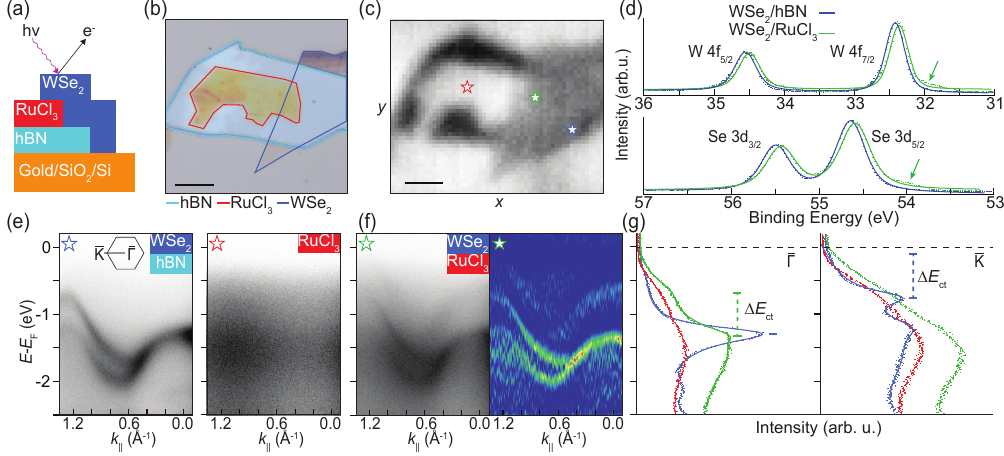}
\caption{Electronic structure of a SL WSe$_2$/RuCl$_3$/hBN heterostructure. (a) Diagram of sample and measurement configuration. (b) Optical image of the sample with hBN, RuCl$_3$ and WSe$_2$ flakes indicated by light blue,  red and dark blue outlines, respectively. (c) Map of ($x$,$y$)-dependent ARPES intensity integrated over momentum from $-0.1$ to $0.1$ \AA$^{-1}$ and energy from $-2.5$ to $-3.5$ eV.  The scale bar in (b)-(c) is 10 µm. (d) Core level spectra (markers) with Voigt fits (lines) from WSe$_2$/hBN and WSe$_2$/RuCl$_3$ areas demarcated by blue and green stars in (c),  respectively.  Arrows indicate a shoulder at lower binding energies.  The spectra were acquired at  $h\nu=145$ eV. (e) Valence band spectra of WSe$_2$/hBN and bare RuCl$_3$ along the $\bar{\mathrm{\Gamma}}-\bar{\mathrm{K}}$ direction of WSe$_2$ from the areas marked by blue and red stars in (c),  respectively.  (f) Valence band ARPES intensity (left) and curvature \cite{zhangPreciseMethodVisualizing2011} (right) of WSe$_2$/RuCl$_3$ from the area marked by a green star in (c).  (g) EDCs (markers) with fits to Lorentzian functions on a linear background (lines) at $\bar{\mathrm{\Gamma}}$ and $\bar{\mathrm{K}}$ from the spectra in (e) and (f). $\mathrm{\Delta}E_{\mathrm{ct}}$ indicates the energy shift of the WSe$_2$ valence band extracted at $\bar{\mathrm{\Gamma}}$.  Red,  blue and green markers and lines demarcate spectra obtained from the areas marked by red,  blue and green stars in (c).  The ARPES spectra were obtained at $h\nu=60$ eV.}
\label{Fig:3}
\end{figure*}

We now turn to electronic structure measurements of SL WSe$_2$/RuCl$_3$ heterostructures.  Due to the sensitivity of RuCl$_3$ towards heating,  we directly transferred the sample from the inert glovebox environment to the microARPES UHV system via a vacuum suitcase after dissolving the PC on PDMS stamp in the glovebox without further heating cycles.  The composition of the sample is illustrated in Fig.  \ref{Fig:3}(a).  An optical micrograph of the sample is presented in Fig.  \ref{Fig:3}(b).  The bulk RuCl$_3$ flake is situated on hBN.  A SL WSe$_2$ flake is placed across the gold-coated substrate and straddles the hBN and a small piece of the RuCl$_3$,  thereby encapsulating this part of the crystal while the remaining part is exposed to vacuum.  This configuration ensures that photo-induced charging is avoided because the top surface is grounded via the SL WSe$_2$ contact with the gold.

The relevant sample locations are found in our microARPES experiment by spatially mapping the ARPES intensity across the area with the flakes.  By integrating the $(E,k)$-dependent photoemission intensity over a spectral region with ranges $-0.1$ to $0.1$ \AA$^{-1}$ and $-3.5$ to $-2.5$ eV and projecting the intensity onto the scanned $(x,y)$-region of the sample,  we produce a spatial map with clear contrasts between the different materials as shown in Fig.  \ref{Fig:3}(c).  A comparison with the optical micrograph in Fig.  \ref{Fig:3}(b) permits us to single-out three distinct spots for further analysis,  which represent bare RuCl$_3$ on hBN (red star),  WSe$_2$/RuCl$_3$ on hBN (green star) and WSe$_2$/hBN (blue star). 

Core level spectra across the W 4f and Se 3d binding energy regions measured in the two distinct WSe$_2$ areas demarcated by green and blue stars are presented in Fig.  \ref{Fig:3}(d).  In both cases we observe the expected doublet of spin-orbit split core levels \cite{chiuDeterminationBandAlignment2015a}, although a slight shift of the main peaks of 0.06 eV is determined via Voigt function fits.  Such a shift can be explained by variations in potential across the  SL WSe$_2$ flake due to charge impurities,  which likely form at interfaces with RuCl$_3$ and hBN,  as these are insulating materials \cite{Ulstrupim:2019,ulstrupSpatiallyResolvedElectronic2016}.  Interestingly,  on the WSe$_2$/RuCl$_3$ spot a broad shoulder is seen towards the low-binding energy side of both the Se 3d$_{5/2}$ and W 4f$_{7/2}$ core levels, as highlighted by a green arrow.  While the shoulder is not possible to reliably fit, we estimate that its centroid is shifted 0.6-0.7 eV from the main core level peak. 

To gain further insights in the electronic structure differences between the three distinct regions we investigate their valence band spectra,  which are shown in Figs.  \ref{Fig:3}(e)-(f).  In all cases the spectra have been collected along the $\bar{\mathrm{\Gamma}}$-$\bar{\mathrm{K}}$ direction of SL WSe$_2$ BZ to track the VBM dispersion and energy shift,  as sketched in Fig.  \ref{Fig:1}(c).  EDCs extracted at the $k_{\parallel}$-values corresponding to $\bar{\mathrm{K}}$ and $\bar{\mathrm{\Gamma}}$ are presented for WSe$_2$/hBN (blue), bare RuCl$_3$ (red) and WSe$_2$/RuCl$_3$ (green) in Fig.  \ref{Fig:1}(g) with fits to multiple Lorentzian functions on a linear background. Fitted EDC peak positions are taken as band positions at the given value of $k_{\parallel}$.

The electronic structure of SL WSe$_2$ on hBN, as shown to the left in Fig.  \ref{Fig:3}(e), exhibits a VBM at $\bar{\mathrm{K}}$ at an energy of -0.77 eV,  which is split by $\mathrm{\Delta}E_{\mathrm{SOC}} = 0.45$ eV.  The energy difference between the VBM at $\bar{\mathrm{K}}$ and the local maximum at $\bar{\mathrm{\Gamma}}$ is determined as $E_{\bar{\mathrm{K}} \bar{\mathrm{\Gamma}}}=0.53$ eV.  The dispersion of SL WSe$_2$ on hBN in our heterostructure is consistent with previous ARPES results on WSe$_2$/hBN \cite{nguyen2019visualizing,wilsonDeterminationBandOffsets2017}.  

The ARPES intensity of the bare part of the RuCl$_3$ flake is shown to the right in Fig.  \ref{Fig:3}(e).  It is  characterized by a flat and broad band centred at -1.5 eV  in addition to a high background.  The flat band is interpreted as the Ru 4d-derived $t_{\mathrm{2g}}$ state discussed in connection with the cleaved bulk RuCl$_3$ flake.  We could not observe the faint Cl 3p-derived bands at lower energies (region not shown) due to the high background intensity, which may be explained by polymer and solvent residues.  The RuCl$_3$ appears reasonably intact after the brief 180$^{\circ}$C heating cycle to melt the polymers,  as the $t_{\mathrm{2g}}$ band is still present and a Fermi level is not established, in contrast to the situation in Fig.  \ref{Fig:2}(d).

In the case of WSe$_2$/RuCl$_3$ in the left panel of Fig.   \ref{Fig:3}(f) two dispersing bands which are rigidly shifted in energy are observed.  From the EDC analysis in Fig.   \ref{Fig:3}(g),  we determine that the dispersion at lower energy aligns with the WSe$_2$ dispersion on hBN shown in Fig.   \ref{Fig:3}(e).  Furthermore,  the intensity level of the upper band is merely 8\% of the level in the lower band.    To better visualize the faint upper band we apply the curvature method to the ARPES intensity \cite{zhangPreciseMethodVisualizing2011}.  The result is presented in the right panel of Fig.   \ref{Fig:3}(f).  Based on the strong resemblance to the WSe$_2$ valence band at lower energy,  we interpret the feature as a rigidly shifted WSe$_2$ valence band due to charge transfer from areas with an intact WSe$_2$/RuCl$_3$ interface.  The lower band is interpreted as stemming from areas where the interfacial contact has been impeded such that the WSe$_2$ valence band adheres to its nominal undoped position as found in WSe$_2$/hBN.  The shift between these two sets of bands thus provides a direct measure of the charge-transfer induced energy shift,  which we determine from the EDCs at $\bar{\mathrm{\Gamma}}$  in Fig.   \ref{Fig:3}(g) to be $\mathrm{\Delta}E_{\mathrm{ct}} = (0.68\pm0.05)$ eV.  At $\bar{\mathrm{K}}$, the shifted band is too faint to reliably fit.  Assuming a rigid shift,  we instead estimate the energy of the VBM by adding $\mathrm{\Delta}E_{\mathrm{ct}}$ to the fitted VBM energy in WSe$_2$/hBN. This places the VBM around -0.09 eV and thereby sufficiently close to $E_{\mathrm{\mathrm{F}}}$ that the VBM filling could be further tuned if this heterostructure was integrated in a gated device geometry \cite{nguyen2019visualizing,Sahoo:2025}. 

The faint shoulder observed in the core level spectra of WSe$_2$/RuCl$_3$ in Fig.   \ref{Fig:3}(d) is consistent with the shifted SL WSe$_2$ valence band and is therefore also attributed to a charge-transfer induced rigid shift.  The inhomogeneous charge-transfer effect pertaining to only 8\% of the sample within the probed 4 µm area may be explained by an incomplete encapsulation of RuCl$_3$,  as SL TMDs develop nano- and microscopic cracks and pinholes during transfer  \cite{Ulstrupim:2019}.  The RuCl$_3$ may additionally be partially compromised by exposure to polymers and solvents creeping under the SL WSe$_2$ as well as the heating cycles to melt the polymers.  We found the RuCl$_3$ clearly decomposes following additional heating steps to clean the sample, as well as extended exposure to solvents during the fabrication process.  When making such changes to the fabrication and preparation procedures, it was not possible to reproduce the charge-transfer effect in other WSe$_2$/RuCl$_3$ samples.

In conclusion, we have used microARPES to directly measure a charge-transfer induced energy shift of $0.68\pm0.05$ eV in SL WSe$_2$ on RuCl$_3$,  placing the VBM of SL WSe$_2$ near the Fermi level.  We found that RuCl$_3$ degrades upon prolonged annealing to 200$^{\circ}$C and that the SL WSe$_2$/RuCl$_3$ interface is spoiled during dry transfer heterostructure fabrication if not carefully encapsulated.  Intact SL WSe$_2$/RuCl$_3$ interfaces unlock new possibilities to tune the filling of the spin-orbit split valence band of SL WSe$_2$ due to the large charge-transfer energy shift,  making ohmic contacts possible.  The insulating RuCl$_3$ does not screen Coulomb interactions,  in contrast to charge-transfer contacts based on metals.  The RuCl$_3$ contact is therefore a promising avenue for establishing,  tuning and spectroscopically probing correlated phases involving the valence band of SL and twisted bilayer WSe$_2$.

\section{Acknowledgements}
The work was funded/co-funded by the European Union (ERC grant EXCITE with project number 101124619). Views and opinions expressed are
however those of the author(s) only and do not necessarily reflect those of the European Union or the European
Research Council.  Neither the European Union nor the granting authority can be held responsible for them.  The authors acknowledge funding from the Novo Nordisk Foundation (Project Grant NNF22OC0079960).  C.S. acknowledges the Marie Sklodowska-Curie Postdoctoral Fellowship (proposal number 101059528). K.W. and T.T. acknowledge support from the CREST (JPMJCR24A5), JST and World Premier International Research Center Initiative (WPI), MEXT, Japan.

\section{Data Availability}
The data used in this study is available on Zenodo \cite{Zenodo}.

\end{document}